## Title

Self-Guided Virtual Reality Therapy for Anxiety: A Systematic Review

## Author Information

Name: Winona Graham
Affiliation: School of Psychology, Charles Sturt University, Bathurst, NSW, Australia.
Email: wgraham@csu.edu.au
ORCID: 0009-0003-0787-750X

Name: Russell Drinkwater
Affiliation: School of Psychology, Charles Sturt University, Bathurst, NSW, Australia.
Email: info@russelldrinkwater.com
ORCID: 0009-0000-7066-4206

Name: Joshua Kelson*
Affiliation: School of Psychology, Charles Sturt University, Bathurst, NSW, Australia.
Email: jkelson@csu.edu.au
ORCID: 0000-0002-3755-175X

Name: Muhammad Ashad Kabir*
Affiliation: School of Computing Mathematics and Engineering, Charles Sturt University, Bathurst, NSW, Australia.
Email: akabir@csu.edu.au
ORCID: 0000-0002-6798-6535

*Corresponding authors

## Abstract

**Background.** Virtual reality (VR) technology can be used to treat anxiety symptoms and disorders. However, most VR interventions for anxiety have been therapist guided rather than self-guided.

**Objective**. This systematic review aimed to examine the effectiveness and user experience (i.e., usability, acceptability, safety, and attrition rates) of self-guided VR therapy interventions in people with any anxiety condition as well as provide future research directions.

**Method**. Peer-reviewed journal articles reporting on self-guided VR interventions for anxiety were sought from the Cochrane Library, IEEE Explore Digital Library, PsycINFO, PubMED, Scopus, and Web of Science databases. Study data from the eligible articles were extracted, tabulated, and addressed with a narrative synthesis.

**Results**. A total of 21 articles met the inclusion criteria. The findings revealed that self-guided VR interventions for anxiety can provide an effective treatment of social anxiety disorder, public speaking anxiety, and specific phobias. User experiences outcomes of safety, usability, and acceptability were generally positive and the average attrition rate was low. However, there was a lack of standardised assessments to measure user experiences.

**Conclusion**. Self-guided VR for anxiety can provide an engaging approach for effectively and safely treating common anxiety conditions. Nevertheless, more experimental studies are required to examine their use in underrepresented anxiety populations, their long-term treatment effects beyond 12 months, and compare their effectiveness against other self-help interventions for anxiety (e.g., internet interventions and bibliotherapy).

## 1 Introduction

Anxiety is an emotional response to perceived threats of future danger (American Psychiatric Association [APA], 2022). Signs and symptoms of anxiety include worries about danger (e.g., being hurt, losing control, or dying), tense muscles, physical trembling, rapid breathing, sweating, nausea, difficulty concentrating, sleep disturbance, and escape and avoidance behaviours (APA, 2022). If a person's anxiety is excessive or prolonged, then they may be classified with an anxiety disorder (APA, 2022). These are separation anxiety disorder, selective mutism, specific phobia, social anxiety disorder, panic disorder, panic attack specifier, agoraphobia, generalised anxiety disorder, substance/medication induced anxiety disorder, anxiety disorder due to another medical condition, other specified anxiety disorder, and unspecified anxiety disorder (APA, 2022). The various anxiety disorders are distinct based on certain aspects, such as the types of objects or situations that are feared, the associated thoughts and behaviours, and the typical age at onset (APA, 2022). Treatment is essential as anxiety can impair daily functioning, worsen quality of life, and increase suicide risk (Bandelow et al., 2017, Kanwar et al., 2013; Wilmer et al., 2021).

Virtual reality (VR) interventions can effectively treat anxiety symptoms and disorders (Andersen et al., 2023; Baghaei et al., 2021; Horigome et al., 2020; Schröder et al., 2023). VR involves using computerised devices, such as a head-mounted display (HMD) or Cave Automatic Virtual Environment (CAVE), to deliver immersive virtual worlds (Menzies et al., 2016; Pellas et al., 2020). These worlds are often used for virtual reality exposure therapy (VRET), which involves people digitally confronting and desensitising themselves to anxiety-inducing stimuli and scenarios (e.g., spider encounters, medical operations, or public speaking; Carl et al., 2019; Emmelkamp & Meyerbröker, 2021). VR therapy for anxiety may also include psychoeducation (Pallavicini et al., 2022), relaxation (Veling et al., 2021), cognitive and behavioural skill development (Emmelkamp & Meyerbröker, 2021), and cognitive distraction from distressing experiences (Wang et al., 2022). For instance, VR can reduce anxiety levels by immersing clients into relaxing virtual nature scenes involving beaches, meadows, mountains, and the sea (Riches et al., 2024). VR interventions for anxiety can either be standalone or used in conjunction with other therapies (e.g., cognitive therapy, medication) to complement professional practice (Andersen et al., 2023; Schröder et al., 2023).

VR interventions for anxiety often have therapist guidance (Donker et al., 2019; Shahid et al., 2024). This is where the therapists play an active role in guiding and facilitating the virtual reality experience for the client (McMahon & Boeldt, 2021). The therapist's involvement ensures proper implementation of the intervention, adherence to the treatment protocol, and monitoring of the client's progress and reactions during the session (Jingili et al., 2023). Therapists can exert control over the virtual environment, manipulating stimuli and adjusting the level of exposure to suit the client's needs (Jingili et al., 2023; Rowland et al., 2022). Their expertise in psychotherapy techniques allows them to provide real-time support, guidance, and feedback to help clients navigate challenging situations and develop coping strategies (McMahon & Boeldt, 2021). This approach also allows therapists to build a strong therapeutic alliance with clients, fostering trust, collaboration, and rapport (Horigome et al., 2020). However, there are limitations with therapist support. Therapist guided VR sessions require the physical presence of a trained professional, which can be logistically challenging for people in geographically remote areas or who have limited access to mental health services (Jingili et al., 2023). The involvement of a therapist also increases the overall cost of treatment because it requires dedicated time and expertise. This cost factor poses a barrier to widespread adoption and affordability of VR therapy interventions (Caponnetto et al., 2021). Moreover, the reliance on therapist guidance may create a dependency on the therapist's presence (Clemens, 2010), which could potentially make it challenging for individuals to generalise skills learned in the virtual environment to real-world situations independently. The fidelity of delivery and obtained VR therapy outcomes for clients may also vary depending on the skills and experience of the therapist (McMahon & Boeldt, 2021). Given these issues, a potential workaround is for VR therapy to be self-guided.

Self-guided VR therapy for anxiety involves people engaging in the therapeutic experience without the direct presence and guidance of a therapist (Donker et al., 2018; Jingili et al., 2023). In this approach, individuals have control over their VR sessions, such as the selection of environments and the level of exposure to feared stimuli (Premkumar et al., 2021). Self-guided VR therapy can potentially empower individuals to take an active role in their own treatment and provide them with greater autonomy, flexibility, and convenience (Jingili et al., 2023). It allows individuals to engage in therapy at their own pace and in the comfort of their own space, reducing barriers such as travel time and scheduling constraints (Jingili et al., 2023). Self-guided computerised treatment of anxiety can increase access to mental health interventions, particularly for individuals in remote areas or those who encounter limitations in seeking face-to-face therapy (e.g., long waitlists; Haug et al., 2012; Jingili et al., 2023; Porter et al., 2023). It also promotes self-efficacy and empowers individuals to develop and practice coping skills independently (Jingili et al., 2023).

Despite the potential benefits of self-guided VR therapy for anxiety, there are user experience issues that need consideration. A common health and safety risk with VR is simulator sickness, which involves negative symptoms (e.g., eyestrain, headache, nausea, vertigo, and dizziness) wrought by a visual and sensorimotor mismatch between the user's body movements and the virtual environment (Balk et al., 2013; Cullen et al., 2021). It is possible that self-guided VR treatment may also aggravate pre-existing conditions (e.g., headaches, epileptic seizures), lead to collisions with physical objects, carry a hygiene risk of infection due to hardware being made of fomite material, and heighten psychological distress during VRET that could be poorly managed without therapist support (Cullen et al., 2021; Goldsworthy et al., 2024; Kelson et al., 2021; Wang & Chan, 2024). It is also important to consider if clients find the self-guided VR technology to be usable (i.e., easy to learn, efficient in performing tasks, not prone to errors, memorable, and satisfying to use; Nielsen, 2012) and acceptable (i.e., they are agreeable towards using the system; Cullen et al., 2021). Negative user experiences regarding safety, usability, and acceptability may lead to higher attrition and poorer therapy outcomes (Balk et al., 2013; Benbow & Anderson, 2019; Cullen et al., 2021).

The existing reviews on virtual reality (VR) therapy for anxiety highlight the growing interest in both therapist-guided and self-guided interventions. For instance, a systematic review by Pelucio et al. (2024) compares therapist-guided and self-guided cognitive-behavioral therapy (CBT), emphasizing the potential of self-guided approaches but falling short of focusing exclusively on VR-based methods. Similarly, reviews such as those by Dhunnoo et al. (2024) and Wiebe et al. (2022)

investigate extended reality therapies and VR for mental health diagnostics and treatment, respectively, but often prioritize therapist-guided interventions and lack a detailed examination of self-guided VR applications. Wu et al. (2021) and Andersen et al. (2023) review VR-assisted CBT and broader VR interventions for anxiety disorders, focusing on therapeutic effectiveness but neglecting a thorough analysis of self-guided applications and user experience. Shahid et al. (2024) and Baghaei et al. (2021) focus on user experience and effectiveness in VR therapies for social anxiety and mental health disorders but lack specificity regarding standalone self-guided VR interventions. Furthermore, Oing and Prescott (2018) review the implementations of VR for anxiety-related disorders but provide limited insights into the standalone self-guided approach. To our knowledge, there is currently no review appraising the effectiveness of self-guided VR interventions for anxiety treatment and concomitant user experience outcomes. To address this gap and help inform clinicians, researchers, and clients interested in exploring this mental health domain, this review examines existing literature to answer the following research questions:

RQ 1. Is self-guided VR therapy an effective treatment modality for anxiety?

RQ 2. What user experience outcomes have been reported regarding safety, usability, acceptability, and attrition of self-guided VR therapy interventions for anxiety?

RQ 3. What future research opportunities are available for self-guided VR therapy for anxiety?

This review synthesises evidence on self-guided virtual reality (VR) therapy for anxiety, highlighting its effectiveness in reducing symptoms with small to large effect sizes and its moderate to high usability and acceptability in unsupervised settings, with minimal simulator sickness. It emphasises the engaging potential of interactive features, such as gamification and virtual therapists, while identifying a critical need for research on its effectiveness and user experience in underrepresented anxiety populations. This paper provides a comprehensive overview and outlines key directions for future research.

## 2 Method

This systematic review was conducted using the Preferred Reporting Items for Systematic Reviews and Meta-Analyses Extension for Systematic Reviews (PRISMA) checklist (Page et al., 2021). The protocol of this systematic review was prospectively registered online with PROSPERO on the 12 August 2024 (CRD42024569216).

### 2.1 Eligibility Criteria

This systematic review included peer-reviewed journal articles written in English that reported on a self-guided VR-based intervention study on participants with clinical or subclinical anxiety. The self-guided intervention needed to make use of immersive VR technology (i.e., HMD or CAVE). The study could have any research design (whether quantitative, qualitative, or mixed methods) and any comparator (e.g., they did or did not compare the use of the self-guided VR intervention to other therapist guided VR interventions or non-VR interventions such as waitlist control, psychotherapy, or pharmacotherapy). The study also needed to report on VR effectiveness outcomes on any standardized anxiety measure as well as participant user experience outcomes (i.e., safety, usability, acceptability, or attrition).

### 2.2 Search Strategy

Eligible articles were sought from the below scientific research databases from 1st January 2012 to 18th June 2024: Cochrane Library, IEEE Explore Digital Library, PsycINFO, PubMED, Scopus, and Web of Science. Keywords used to search each database were (“virtual reality” or “VR”) and (“anxiety” or phobia”) and (“remote” or “self-help” or “self-guided” or “self-care” or “self-led”). Search filters of English language and article type (i.e., journal articles and exclude review articles and protocol registrations) were applied where available in each database. Finally, a manual search of the reference lists of each included article was also performed.

### 2.3 Article Selection

Database search results were exported, deduplicated, and then screened on title and abstract information.  Promising articles were short-listed and obtained for full-text appraisal against eligibility criteria. Final study inclusion was agreed upon by all members of the research team.

### 2.4 Data Extraction and Analysis

A standardized coding sheet developed by one reviewer (JK) was used to guide data extraction from the eligible articles. Data items for extraction included: reference source (authors, publication title and year); study design (methodology, comparators, and measurement time points, such as pre-test, post-test, and any longitudinal follow-up); population (country of origin, sample size, and demographics); VR intervention information (name, virtual environments, hardware, content, and treatment details, such as number of VR therapy sessions and their time length); effectiveness results (methodological quality, measure names, outcomes and treatment effect sizes); and participant user experience findings (acceptability, safety, attrition, usability, and intention-to-treat analyses). Attrition in this review was defined and measured as the relative number of participants who started using the self-guided VR intervention at pre-test but did not complete measurements at post-test. Data were extracted and checked by two reviewers (RD, WG) to reduce reviewer bias and correct any potential data extraction errors.

The Mixed Methods Appraisal Tool was used to assess the quality of all included studies (Hong et al., 2018). This critical appraisal tool involves categorising each study based on research design (i.e., randomized controlled trial [RCT], non-randomised quantitative study, quantitative descriptive study, qualitative study, and mixed methods study). Each study has two screening questions and then five questions on methodological quality based on their respective category.

Extracted data from the included studies were tabulated for analysis. As the observed findings involved a mix of quantitative and qualitative data obtained from diverse methodologies, a narrative synthesis approach was adopted for this review rather than statistical meta-analysis (Siddaway et al., 2019). The narrative synthesis approach involves answering the review's research questions using text to summarise and explain the findings of the included studies.

# 3 Results

## 3.1 Study Selection

Figure 1 shows that the literature search yielded 414 records. A total of 244 remained after de-duplicating citations. Of these records, 21 articles met the eligibility criteria after screening based on title, abstract, and full-text appraisal. There were 19 unique studies found on self-guided VR therapy for anxiety. Two articles (Donker et al., 2020; Kahlon et al., 2023b) are additional reports on other included studies in the review.

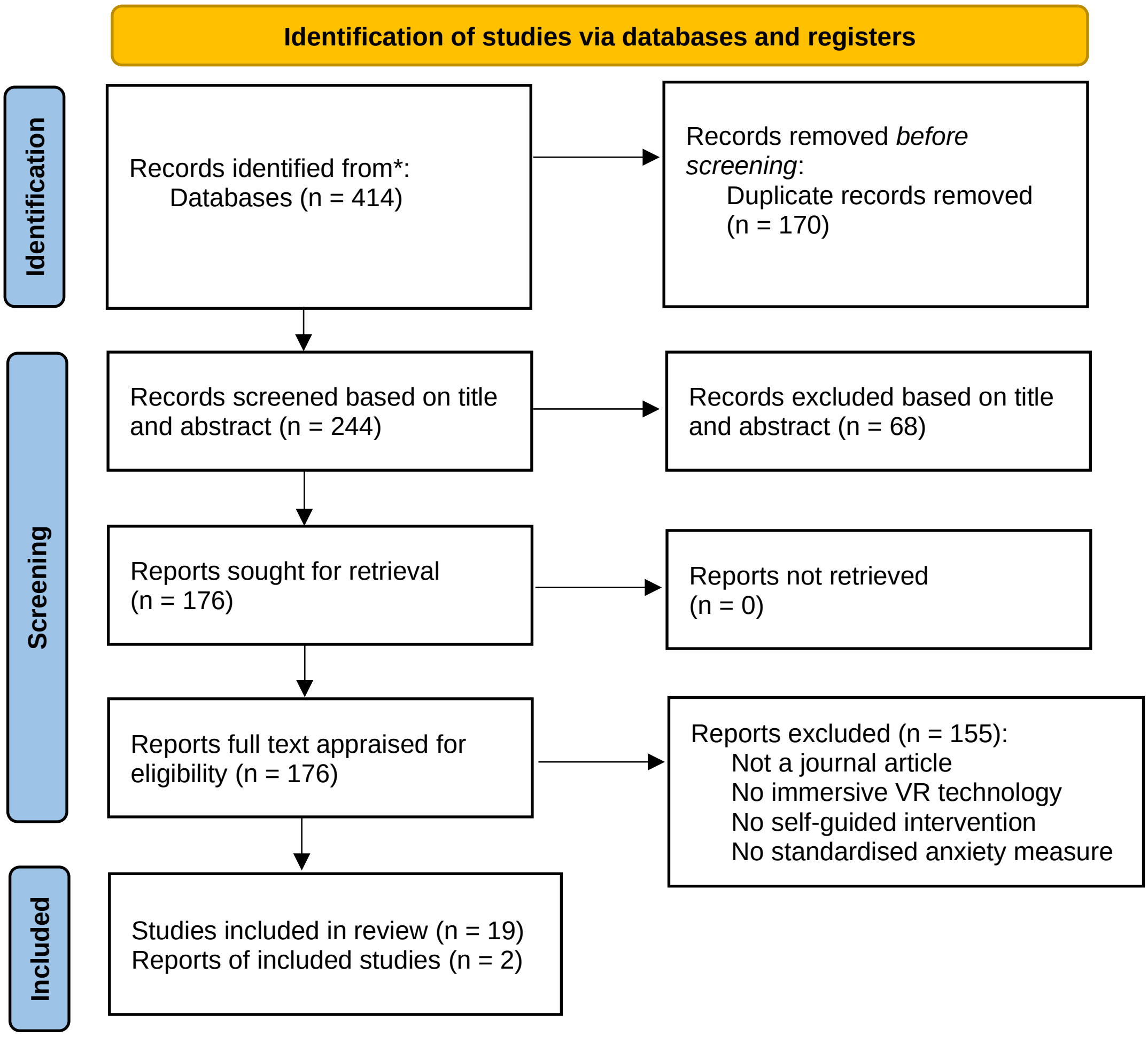

**Fig. 1** Flowchart of systematic review search results.

### 3.2 Participant Characteristics

Participants details across the included studies can be seen in Table 1. Participants were recruited from twelve countries of origin, with the largest proportion of participants from the Netherlands (*N* = 243), followed by New Zealand (*N* = 126), Norway (*N* = 100), the USA (*N* = 98) and UK (*N* = 80). Other countries with smaller representations include Sweden (*N* = 75), Iran (*N* = 70), Japan (*N* = 70), South Korea (*N* = 54), Germany (*N* = 38), Italy (*N* = 40), and Canada (*N* = 6). Sample sizes ranged from 6 to 193 participants, with a median of 40 participants. Participant ages ranged from 12 – 67 years. Participants were mainly female with an average sample proportion of 69.76% (range 24% to 90%). Both clinical (76.9%, *N* = 769) and non-clinical (23.1%, *N* = 231) studies were conducted. Most participants had anxiety, for example, public speaking anxiety, social anxiety disorder, a specific phobia, or panic disorder. However, some studies focused on other mental illnesses, such as depression, bipolar disorder, psychotic disorder, PTSD, or chronic pain, and only measured anxiety as a secondary outcome. In clinical studies, diagnoses were obtained by questionnaires or clinical interviews delivered in person, by phone, or videoconferencing. In contrast, non-clinical studies focused on stress, fear related to COVID-19, varied measures of psychological distress or negative emotional states. Two non-clinical studies (*N* = 100 participants) specified a major mental illness as an exclusion criteria.

Table 1. Description of Participants and Research Designs in Reviewed Studies

| *Study* | *Country* | *Problem/diagnosis* | *N*[a] | *Mean age (SD*[b]*) in years* | *Age range in years* | *n (%)*[c] | *Design* | *Treatment conditions (n)*[d] | *Measurements* |
|---|---|---|---|---|---|---|---|---|---|
| Chard et al. (2023) | UK | Social anxiety | 25 | VRET[e]: 32 (9.44)<br><br>Waitlist: 39 (16.86) | - | 24 | RCT[f] | VRET (13); Waitlist (12). | Pre-test, post-test, and 1-month follow-up |
| Donker et al. (2019; 2020) | The Netherlands | Acrophobia | 193 | 41.33 (13.64) | 18-65 | 66.84 | RCT | VR group (96); Waitlist (97) | Pre-test, post-test, and 3-month follow-up |
| Farahimanesh et al. (2023) | Iran | Non-clinical distress and fear of COVID-19 | 60 | Intervention: 49.1 (10.92)<br><br>Control: 49.7 (10.4) | - | 55 | RCT | VR group (30); Control group (30) | Pre-test, post-test, and 2-week follow-up |
| Kahlon et al. (2023a; 2023b) | Norway | Public speaking anxiety | 100 | 14.2 (0.99) | 13-16 | 84 | RCT | VR only (20); VR + online exposure program (20); Online psychoeducation + online exposure program (40); Waitlist + online psychoeducation program (20) | Pre-test, during-test, post-test, and 3-month follow-up |
| Khaleghi et al. (2024) | Iran | Ailurophobia | 10 | Non biofeedback VRET: 24 (7.31); Biofeedback VRET: 33.5 (7.16) | - | 90 | RCT | Biofeedback VRET (5); non-biofeedback VRET (5) | Pre-test and post-test |
| Lacey et al. (2023 ) | New Zealand | Specific phobia (flying, heights, spiders, dogs and needles) | 126 | 42.2 (13.2) | 18-64 | 80 | RCT | VR group (63); Waitlist (63) | Pre-test, post-test, and 12-week follow-up |
| Lindner et al. (2020) | Sweden | Fear of spiders | 25 | 25(11) | - | 76 | Quantitative nonrandomised | VRET group (25) | Pre-test, post-test, and follow-ups at 1-week, 2-week, and 6-months |
| Lindner et al. (2019) | Sweden | Public speaking anxiety | 50 | Therapist-led VRET: 30.84 (6.63)<br><br>Waitlist/self-led VRET: 31.88 (7.91) | - | Therapist-led VRET: 80<br><br>Waitlist/self-led VRET: 64 | RCT | Therapist-led VRET (25)<br><br>Waitlist and self-led VRET (25) | Pre-test, during-test, and post-test |
| Matsumoto et al. (2021) | Japan | Non-clinical stress | 70 | - | - | - | Quantitative nonrandomised | VR app (24) CB[g] app (23) Combined VR app + CB app (23) | Pre-test and post-test |
| Meyer et al. (2022) | Germany | Non-clinical distress and fear of COVID-19 | 38 | 36.4 (12.5) | 20-67 | 73.7 | Quantitative nonrandomised | VR Treatment (38) | Pre-test, post-test, and 2-week follow-up |
| Miller et al. (2023) | USA | Depression | 30 | 17.03 (2.97) | 12-21 | 70 | Mixed Methods | Spark CBT[h] app and VR (30) | Pre-test, Post-test, and 1-month follow-up |
| Piercey et al. (2012) | Canada | Spider phobia | 6 | - | - | - | Quantitative nonrandomised | VRET (6) | Pre-test, during-test, and post-test. |
| Premkumar et al. (2021) | UK | Public Speaking anxiety | 32 | 21.4 (4.9) | 18-40 | 84.4 | Quantitative nonrandomised | VRET (32) | Pre-test, during-test, post-test, and 1-month follow-up |
| Riches et al. (2024) | UK | Non-clinical stress and anxiety | 23 | 22.96 (2.21) | 21-29 | 65.2 | Mixed Methods | VR treatment (11); Control (12) | Pretest and post-test |
| Riva et al. (2021) | Italy | Non-clinical distress and fear of COVID-19 | 40 | 30.28 (11.69) | - | 62.5 | Quantitative nonrandomised | VR Treatment (40) | Pre-test, during-test, post-test, and 2-week follow-up |
| Senda et al. (2023) | USA | Dysphoria (depression, anxiety, PTSD, or chronic pain) | 24 | 43.6 (15.8) | - | 63 | Quantitative nonrandomised | VR guided mindfulness (23). TMS[i] left dlPFC (19) TMS dmPFC (13) | Pre-test, during-test, post-test, and follow-up |
| Shin et al. (2021) | South Korea | Panic disorder | 54 | - | 19 - 60 | VR group: 60; Waitlist group: 67. | RCT | VR group: (33); Waitlist group (21). | Pre-test and post-test |
| Veling et al. (2021) | The Netherlands | Anxiety, psychotic, depressive, or bipolar disorder | 50 | 41.6 (14.2) | - | 66 | RCT | VR group (25); Relaxation exercises (25) | Pre-test, during-test, and post-test |
| Zainal et al. (2021) | USA | Social anxiety disorder | 44 | 23.30 (9.32) | 18-53 | 77.3 | RCT | VRE group (26); Waitlist (18) | Pre-test, post-test, and follow-ups (3- and 6-months) |

[a] Refers to the total number of participants in the study
[b] SD: standard deviation
[c] Refers to the percentage of female participants
[d] Refers to the number of participants in each treatment condition
[e] VRET: virtual reality exposure therapy
[f] RCT: randomized control trial
[g] CB: chatbot
[f] CBT: cognitive behavioural therapy
[i] TMS: transcranial magnetic stimulation

### 3.3 Research Designs and Comparators

Table 1 shows that participant VR use was appraised in quantitative RCT studies (*n* = 10, 52.63%), quantitative nonrandomized studies (*n* = 7, 36.84%), and mixed methods studies (*n* = 2, 10.53%). Seven studies used a single-group research design. To collect data on VR outcomes, all studies used surveys or scales, four studies used semi-structured interviews (Khaleghi et al., 2024; Meyer et al., 2022; Miller et al., 2023; Riches et al., 2024), four studies obtained direct user commentary (Chard et al., 2023; Lacey et al., 2022; Piercey et al., 2012; Zainal et al., 2021), and three studies collected user engagement data during experimentation (Donker et al., 2020; Miller et al., 2023; Veling et al., 2021). Comparators included waitlist, no treatment control group, an online exposure program, online psychoeducation, therapist-led VR, cognitive behavioural therapy app, and relaxation exercise. All studies had pre-test and post-test assessments on user outcomes, though 14 also had follow-up assessments, and seven had assessments during the active VR test phase.

### 3.4 Details of the Virtual Reality Interventions

Table 2 shows that most studies tested a unique VR intervention, except for studies that conducted the same research in Iran (Farahimanesh et al., 2023), Germany (Meyer et al., 2022), and Italy (Riva et al., 2023). Hardware involved a standard computer, smartphone, and HMDs, which ranged from inexpensive cardboard VR viewers to more expensive headsets such as Oculus Quest 1, Samsung Gear VR headset, or a Pico Interactive Goblin VR headset. The software involved passive or active virtual environments with audio, video, text, interactivity, and educational content. The passive virtual environments entailed watching relaxing or pleasant videos such as virtual gardens, beaches, mountains, or ocean scenes. Active virtual environments typically involve graded exposure to treat public speaking anxiety, social anxiety disorder, or a broad range of phobias including fear of cats, dogs, spiders, heights, flying, and needles. Treatment lengths ranged from a single session to five weeks, with a mode of three weeks of VR. Participant VR usage time ranged from two minutes to three hours per session or module, and two studies delivered the VR in a single session.

Table 2. Details of the Virtual Reality Interventions

| Study | Virtual Environments | Headset | Treatment Length (duration) |
|---|---|---|---|
| Chard et al. (2023) | Ordering a drink at a café, using a telephone, public speaking. | Virtual Real Store Google Cardboard V2 headsets | Three once weekly sessions. Exposure sessions, completed at their own pace. |
| Donker et al. (2019; 2020) | Changing a light bulb on a small ladder, connecting speakers at the edge of the stage, going up a high ladder to repair a small, damaged platform, fixing a spotlight on the highest balcony, saving a cat while being on a gangway high above the stage. | Cardboard VR viewer | Three weeks. Six animated modules of the VR-CBT app and exposure (5-40 mins each). |
| Farahimanesh et al. (2023) | Relaxing video titled "Secret Garden" and social tasks with specified objectives. | Cardboard VR headset | Daily for one week (20 mins each day) |
| Kahlon et al. (2023a; 2023b) | Public speaking exposure with four adjustable variables: number of people in the audience, | Oculus Quest 1 | VR only. Participants were instructed to complete at |
| | duration of task, audience reaction (uninterested, neutral, or interested), and type of presentation (sitting by their desk and reading from a book or presenting in front of the classroom). | | least five tasks a week and practice three times a week. One task took five minutes. Weekly time commitment up to 60 minutes.<br><br>VR + online exposure program. Participants completed the VR component in the first three weeks, doing one module per week. Then weeks four to six were spent practicing three in vivo exposure tasks per week. |
| Khaleghi et al. (2024) | Varying cat exposure e.g. cat photos, fantasy cat models, low-poly cat models with minimal details, and high-poly cat models that closely resemble real cats. Cats could either be still, shaking their head, cleaning their paws, or following the participant. They could be silent or making noise. | Brand Not Reported | Four mandatory game levels, with each session lasting up to three hours. |
| Lacey et al. (2022) | Scenes exposing participants to phobic stimuli such as fear of flying, heights, spiders, dogs and needles. | Brand Not Reported | VR intervention consisted of six modules. The first four modules (psychoeducation, relaxation, mindfulness, and cognitive techniques) were completed in two weeks. The VR exposure therapy and relapse prevention modules were to be completed over the remaining four weeks (5–10 minutes each day). |
| Lindner et al. (2020) | Varying exposure to spiders. Included eight gamified levels with gaze focusing and gaze directed tasks. Participants also helped a spider complete tasks. | Samsung Galaxy S6 and Gear VR headset | A single session of three hours. |
| Lindner et al. (2019) | Participants deliver a speech in an auditorium, wedding reception, or meeting room. | Samsung Gear VR headset (1st generation) (Therapist led); Google Cardboard headset (Self-led). | A single session of three hours. After VR exposure, the in vivo transition program consisted of four modules, one released for completion each week |
| Matsumoto et al. (2021) | Visual stimulation of viewing pleasant images such as lights and smiles. | Samsung GearVR | Four weeks. The VR group had no message to encourage its use.<br>The CB group and VR + CB group received daily messages to use VR to release stress |
| Meyer et al. (2022) | Relaxing video titled "Secret Garden" and social tasks with specified objectives. | Cardboard VR Basetech Headmount Google 3D | Daily for one week (20 mins each day) |
| Miller et al. (2023) | Psychoeducation, mindful breathing, and guided meditations | Pico Interactive Goblin VR headset, | Five weeks. One module to be completed every week. |
| Piercey et al. (2012) | Exposure to spiders in a virtual apartment setting across five levels. | Red/blue anaglyph glasses | Four treatment sessions of 25 minutes. |
| Premkumar et al. (2021) | Delivering a speech in a virtual classroom. Participants could vary the audience size and reaction as well as their number of props, salience in the room, and distance from the audience. | Samsung Gear VR headset | Two treatment sessions over two weeks (60 mins each). |
| Riches et al. (2024) | VR relaxation scenes e.g. virtual beaches, mountains, meadows and the sea. | Oculus Go Wireless HMD | Fourteen days, encouraged to do at least one VR session per day. Participants chose the time and duration of VR. |
| Riva et al. (2021) | Relaxing video titled "Secret Garden" and social tasks with specified objectives. | Cardboard VR headset | Daily for one week (20 mins each day) |
| Senda et al. (2023) | Mindfulness exercises. | Valve Index Headset | Two weeks. Two x 15-minute VR sessions per day with a 50-minute break in between. |
| Shin et al. (2021) | Driving a car, catching a plane, riding in an elevator or on the subway. | Samsung Gear VR | Three times per week, for four weeks. There were 12 sessions, lasting 15-30 mins each. |
| Veling et al. (2021) | Relaxing scenes including beaches, a coral reef, swimming underwater with wild dolphins, mountain scenery in the Alps, drone flights and cliffside views. | Samsung Gear VR | Daily for 10 days (10+ mins a day) per intervention, VR and relaxation. |
| Zainal et al. (2021) | An informal dinner party or formal job interview. | Pico Goblin VR headset | Lab visits twice a week for 50-60 minutes, including 25-30 minutes of VR immersion. Expected to participate in four to 10 sessions and in vivo homework. |

VR: virtual reality
CBT: cognitive behavioural therapy
CB: chatbot

### 3.5 Effectiveness Measures and Outcomes

Details of the measures and outcomes of VR treatment on anxiety are summarised in Table 3. Most studies (*n* = 16, 76.19%) reported a significant reduction from pre-test to post-test VR-based treatment on standardised anxiety measure. VR treatment effect sizes across all studies that reported them ranged from small to large in magnitude.

Table 3. Details on Anxiety Measures and Reported Outcomes

| Study | Measures | VR effectiveness outcomes |
|---|---|---|
| Chard et al. (2023) | Social Phobia Scale (SPS) | There was a slight reduction in social anxiety for the VRET group from pre- to post-intervention and an increase in social anxiety for waitlist controls. However, the difference was not significant ($p$ = .09). |
| | Fear of Negative Evaluation Scale (FNE-B) | The FNE-B scores remained stable from pre- to post-test with no significant difference between the VRET and waitlist groups ($p$ = .19). |
| | Unhelpful Thoughts and Beliefs About Stuttering scale (UTBAS-6) | The UTBAS-6 scores remained stable in the VRET group from pre- to post-intervention and while there appeared to be a small decline for the waitlist controls ($d$ = 0.25), the difference between groups was not significant ($p$ = .59). |
| Donker et al. (2019, 2020) | Acrophobia Questionnaire (AQ) | A significant decrease in acrophobia symptoms was found for the VR group compared to waitlist controls post-intervention (*p* < .001), with a large effect size (*d* = 1.14). For participants who completed the AQ assessment post-intervention, there was a large between-group effect size (*d* = 1.53). The within-group effect size between pre-test and 3-month follow-up was large (*d* = 2.68). More practice time in VR sessions was significantly associated with lower acrophobia symptoms post-intervention (*p* < .05). |
| | Attitudes Toward Heights Questionnaire (ATHQ) | The ITT analysis showed a significant decrease in ATHQ scores for the VR group compared to waitlist controls from pre- to post-intervention with a large effect size (*p* < .001; *d* = 1.09) which was maintained at 3-month follow up. |
| | Beck Anxiety Inventory (BAI) | The ITT analysis showed a significant decrease in anxiety for the VR group compared to waitlist controls from pre- to post-intervention with a medium effect size (*p* < .001; *d* = 0.37) which was maintained at 3-month follow up. |
| Farahimanesh et al. (2023) | Depression Anxiety Stress Scale (DASS-21) | The anxiety subscale showed a significant main effect of time (*p* < .001) and group (*p* < .001) and an interaction effect time × group (*p* < .001). Post-hoc comparisons revealed significant reductions in anxiety from pre- to post-intervention for participants in the VR group (p < .001) that were maintained at 2-week follow-up. |
| | Fear of Coronavirus (FCOR) | Participants in the VR group showed a significant reduction in fear of COVID-19 from pre- to post-intervention (*p* < .001), which remained stable at two weeks follow up. |
| Kahlon et al. (2023a, 2023b) | Public Speaking Anxiety Scale (PSAS) | There was a significant decrease in public speaking anxiety symptoms in the VR only group from pre- to post-intervention (*p* = .026). The linear mixed model analysis indicated a significant decrease in public speaking anxiety for the combined VRET groups, with an average decrease of 2.14 points per week over the 3-week intervention (*p* < .001) and a large within-group effect size (*d* = 0.83). The VRET groups showed significantly larger reductions in PSAS scores than the waitlist group (*p* < .015), with a moderate between effect size (*d* = 0.61). |
| | Social Phobia Scale (SPS-6) | There was a significant decrease in social phobia symptoms in the VR only group from pre- to post-intervention (*p* = .035). The linear mixed model analysis showed a significant decrease in social phobia for the combined VRET groups, with a reduction of 1.91 points each week during the intervention phase (*p* = .007) and a small within-group effect size (*d* = 0.22). There was also a significant decrease in social phobia symptoms for the VR + online exposure group, with a reduction of 3.11 points during intervention (*p* < .009) and a small within-groups effect size (*d* = 0.36). There was also a significant decrease in social phobia symptoms in the online psychoeducation + exposure group from pre- to post-intervention (*p* = .023). |
| | Social Interaction Anxiety Scale (SIAS-6) | There was no significant change in social interaction symptoms for any group from pre- to post-intervention. The only groups that had a significant decrease in social anxiety symptoms from post-intervention to follow-up were waitlist + online psychoeducation (*p* = .043) and online psychoeducation + exposure (*p* = .041). |
| Khaleghi et al. (2024) | State-Trait Anxiety Inventory (STAI) | There was a 50-point increase in state anxiety for the VRET non-biofeedback group and a 33-point increase for the VRET biofeedback group, with no significant difference between the groups post-intervention (*p* > .05). |
| | Fear of Cats Questionnaire (FCQ) | The VRET non-biofeedback group showed a 67-point improvement in FCQ scores, while the biofeedback group deteriorated by 42 points, yet there was no significant difference between the groups post-intervention (*p* > .05). |
| Lacey et al. (2023) | Severity Measures for Specific Phobia – Adults (SMSP) | There was a significant decrease in phobia symptoms from pre-intervention to week 6 post-intervention in the VR group compared to the waitlist group (*p* < .001), with a large effect size of 0.86 reported. The VR group maintained their treatment gains at follow-up. |
| | Brief Fear of Negative Evaluation Scale (BFNE) | The VR group demonstrated a small reduction in fear of negative evaluation from pre-intervention to week 6 post-intervention, while the waitlist group demonstrated no significant change (*p* >.07). |
| Lindner et al. (2020) | Fear of Spiders Questionnaire (FSQ) | Mixed effects modelling revealed a significant reduction in the fear of spiders after VRET (*p* < .001) with a large within-group effect size of *d* = 1.26. Eight participants (35%) achieved clinically significant change. |
| Lindner et al. (2019) | Public Speaking Anxiety Scale (PSAS) | There was a significant reduction in public speaking anxiety for both therapist-led and self-led VRET interventions. On average, participants in the therapist-led group experienced a reduction in public speaking anxiety that was 6.90 points greater than the waitlist/self-led group, at each treatment step. Following the therapist-led VRET (or equivalent waiting period), there was a 6.17-point difference between the groups, with a large between-group effect size of *d* = 0.83 and a medium within-group effect size of *d* = 0.77. After the in-vivo transition program (or equivalent waiting period), this difference rose to 12.99 points, with a large between-group effect size of *d* = 1.50 and a large within-group effect size of *d* = 1.67. After the therapist-led VRET and in vivo intervention was complete, the waitlist group commenced self-led VRET. After participants completed the self-led VRET, there was an average reduction of 8.91 points in public speaking anxiety scores, with a large within-group effect size of *d* = 1.38. After completing the in-vivo transition program, the average reduction increased to 15.7 points, with a large within-group effect size of *d* = 1.35. The self-led group showed no significant decrease in anxiety between post-intervention and 6-month follow-up (*p* = .279). |
| | Generalised Anxiety Disorder 7-item (GAD-7) | There were no significant changes in generalised anxiety scores detected during either treatment phase, when examining the therapist-led versus waitlist, or waitlist versus self-led groups. |
| | Brief Fear of Negative Evaluation Scale (BFNE) | There were no significant changes in fear of evaluation detected during either treatment phase, when examining the therapist-led versus waitlist, or the waitlist versus self-led groups. |
| Matsumoto et al. (2021) | State Trait Anxiety Inventory (STAI) | There were no significant changes in the nurse's tendencies to become anxious from pre- to post-intervention in the VR, CB and VR + CB conditions. |
| Meyer et al. (2022) | Depression Anxiety Stress Scales (DASS 21) Anxiety | There was a significant decrease in anxiety across measurement points (two pre assessments, post, and follow up), *p* < .001. Anxiety was significantly lower from pre- to post-intervention (*p* < .05) and from pre-intervention to 2-week follow up (*p* < .01). |
| | Fear of COVID-19 scale (FCV-19S) | A significant reduction in the fear of COVID-19 occurred between the waiting period (7 days before intervention) and the first day of intervention but not as hypothesised from pre- to post-intervention (*p* = .43). Fear of COVID-19 significantly reduced from pre-intervention to 2-week follow up (*p* = .01). |
| | Subjective Units of Distress Scale (SUDS) | There was a significant reduction in subjective distress from pre- to post-intervention ($p$ = .013). |
| Miller et al. (2023) | Generalised Anxiety Disorder 7-item (GAD-7) | There was a reduction in anxiety symptoms from pre- to post-intervention and from pre-intervention to the 1-month follow-up, as reflected in the mean difference scores. |
| Piercey et al. (2012) | Fear of Spiders Questionnaire (FSQ) | There was a significant reduction in the spider phobia symptoms from pre- to post-intervention ($p < .01$). |
| | Spider Beliefs Questionnaire (SBQ) | There was a significant reduction in SBQ beliefs and actions from pre- to post-intervention ($p < .01$). |
| | State-Trait Anxiety Inventory (STAI) | There were no significant changes in state or trait anxiety from pre- to post-intervention ($p > .05$). |
| Premkumar et al. (2021) | Speech Anxiety Thoughts Inventory (SATI) | There was significant improvement in SATI scores at VRET session 1, session 2, and at 1-month follow-up relative to pre-intervention ($p \leq .01$). |
| | Public Speaking Anxiety Scale (PSAS) | There was a significant decrease in public speaking anxiety at VRET session 1, session 2 and at 1 month follow-up ($p \leq .01$). |
| | Personal Report of Confidence as a Speaker (PRCS-SF) | There was a significant increase in confidence in public speaking from VRET session 1 to session 2 (p < 0.001). However, confidence scores declined between session 2 and the 1-month follow-up ($p < .001$). |
| | Liebowitz Social Anxiety Scale (LSAS) | There was a significant reduction on the performance anxiety subscale from pre-intervention and session 2 to follow up ($p < .005$). There were no significant changes on the other subscales including performance avoidance, social anxiety, and social avoidance ($p > .05$). |
| | Brief Fear of Negative Evaluation Revised Scale (BFNE) | There was a significant decrease in fear of negative evaluation between pre-intervention, VRET session 2 and 1-month follow-up ($p$ = .002). |
| | Subjective Units of Distress Scale (SUDS) | There was a significant decrease in SUDS-anxiety over the two sessions ($p < .001$) and SUDS-arousal decreased from pre-intervention to session 2 of VRET ($p$ = .003). Post hoc Bonferroni-corrected pairwise comparisons revealed significantly lower levels of anxiety at the end of each VRET session, compared to ratings made during the first two session pauses ($p \leq .001$). |
| Riches et al. (2024) | Generalised Anxiety Disorder Scale (GAD-7) | Pre-intervention, the mean GAD-7 score was 8.27 (*SD* = 5.24) for the VR group and 7.00 (*SD* = 4.82) for the control group. Post-intervention, the mean GAD-7 score was 6.18 (*SD* = 4.73) for the VR group and 5.42 (*SD* = 3.42) for the control group. There was insufficient power to conduct between group comparisons due to small sample size. |
| | Visual Analogue Scale (VAS) | Participants who did VR ≥ 8 times had a significant reduction in anxiety ($p < .001$), with a medium effect size (Pearson's $r = -.34$). However, participants who did VR ≤ 7 times had no significant change in anxiety ($p = .166$). |
| Riva et al. (2021) | Depression Anxiety Stress Scales (DASS-21) Anxiety | There was no significant change in the anxiety subscale post-intervention ($p > .05$). However, compared to the pre-intervention waiting phase, the VR intervention demonstrated a medium effect size with protective effects. |
| | Fear of Coronavirus (FCOR) | Fear of COVID-19 decreased between the waiting phase and the 2-week follow-up ($p = .003$). However, there was no significant decrease from pre- to post-VR treatment ($p = .412$). |
| | State-Trait Anxiety Inventory (STAI) | There was a significant effect of time on state anxiety, with lower STAI scores reported during the intervention phase ($M = 35.29$, $SD = 7.52$) compared to the pre-intervention waiting phase ($M = 42$, $SD = 9.48$). |
| | Subjective Units of Distress Scale (SUDS) | There was a significant effect of time on subjective distress, with lower SUDS scores reported during the intervention phase ($M = 10.71$, $SD = 13.1$) compared to the pre-intervention waiting phase ($M = 17.62$, $SD = 15.6$). |
| Senda et al. (2023) | Generalised Anxiety Disorder-7 (GAD-7) | There were no significant changes in anxiety scores from pre-test to post-test VR treatment ($p = .635$). |
| | Hamilton Anxiety Rating Scale (HAM-A) | There were no significant changes in anxiety scores from pre-test to post-test VR treatment ($p = .113$). |
| Shin et al. (2021) | Body Sensations Questionnaire (BSQ) | A moderate within-group effect size ($d = 0.57$) was found for BSQ scores in the VR group from pre- to post-intervention. There was no significant difference in BSQ scores between the VR group ($M = 60.55$, $SD = 12.22$) and waitlist control ($M = 61.40$, $SD = 11.03$) post-intervention ($p = 81$). |
| | Panic Disorder Severity Scale (PDSS) | The VR group showed significantly lower panic symptoms post-intervention compared to the control group ($p = .003$). A large within-group effect size ($d = 1.05$) was observed, indicating a substantial reduction in panic symptoms within the VR group. Between-group differences of change were also significant ($p = .02$), supporting the effectiveness of treatment. |
| | State-Trait Anxiety Inventory (STAI) | The VR group had significantly lower state anxiety post-intervention compared to the control group ($p = .04$). There was a large within-group effect size ($d = 0.91$), indicating a substantial reduction in state anxiety within the VR group. Between-group differences of change were also significant ($p = .04$). |
| | Hospital Anxiety and Depression Scale (HADS) | The anxiety subscale showed a moderate effect size ($d = 0.59$), indicating a notable change in the VR group from pre- to post-intervention. |
| | Albany Panic and Phobia Questionnaire (APPQ) | There was no significant difference on APPQ scores between the VR group and waitlist control for treatment completers ($p = .80$). |
| | Anxiety Sensitivity Index (ASI) | There was no significant difference on ASI scores between the VR group and waitlist control for treatment completers ($p = .93$). |
| | Korean Inventory of Social Avoidance and Distress Scale (K-SADS) | There was no significant difference on K-SADS scores between the VR group and waitlist control for treatment completers ($p$ = .19). |
| Veling et al. (2021) | Visual Analogue Scales (VAS) | VRelax significantly reduced negative affective states (including anxiety) to a greater extent than relaxation exercises ($p$ = .04) |
| | Beck Anxiety Inventory (BAI) | Symptoms of anxiety were reduced significantly after VRelax and relaxation exercises. There was no significant difference between the effects of the two groups on anxiety ($p$ = .06). |
| Zainal et al. (2021) | SAD Composite of Social Phobia Diagnostic Questionnaire (SPDQ) and Social Interaction Anxiety Scale (SIAS) | There was a significant decrease in social anxiety symptoms from pre- to post-VRE intervention compared with the waitlist group (Hedge's $g$ = −4.77; $p$ <.001). These gains were maintained at 3 and 6-month follow-ups, as there was no significant change in the within VRE SAD severity composite scores. |
| | Measure of Anxiety in Selection Interviews (MASI) | There was a significant decrease in job interview anxiety from pre- to post-VRE intervention compared with the waitlist group (Hedge's $g$ = −4.17; $p$ <.001). These gains were maintained at 3- and 6-month follow-ups. |
| | Penn State Worry Questionnaire (PSWQ) | There was a significant decrease in PSWQ scores in the VRE group compared to the waitlist controls ($p$ < .001). These gains were maintained, as there were no significant changes at 3 and 6-month follow-ups. |

### 3.6 User Experiences Outcomes

The average attrition rate across studies in the self-guided VR treatment phase from pre-test to post-test was 16.18%, ranging from 0 to 50% (see Table 4). Three studies did not report attrition rates, and 10 studies reported an intention-to-treat analysis. Most studies used non-standardised questions on VR user experience factors of satisfaction, acceptability, user-experience, and user-preferences. Six studies used the standardised Simulator Sickness Questionnaire (Donker et al., 2019; Lindner et al., 2020; Meyer et al., 2022; Shin et al., 2021; Veling et al., 2021; Zainal et al., 2021) and 6 included studies used standardised measures on VR user presence. These measures included the Presence Questionnaire (Piercy et al., 2012), Igroup Presence Questionnaire (Donker et al., 2019;

2020; Zainal et al., 2021), Gatineau Presence Scale (Lindner et al., 2020), and Sense of Presence Scale (Riches et al., 2024).

Table 4. Virtual Reality Interventions and User Experience Outcomes

| Study | Measures | VR User Experience Findings | Attrition (%) | ITT[a] |
|---|---|---|---|---|
| Chard et al. (2023) | Qualitative feedback | Several participants experienced discomfort using cardboard headsets. Several participants reported that the exercises lacked applicability to their personal experience of anxiety. Consequently, one participant did not feel anxious during VR sessions and ceased treatment while another did not feel connected to the virtual therapist. | 4/13 (30.8) | Yes |
| Donker et al. (2019; 2020) | System Usability Scale (SUS), Igroup Presence Questionnaire (IPQ), Simulator Sickness Questionnaire (SSQ) | The VR-CBT app received a user-friendly rating (mean [SD] = 75.35 [14.74], n = 55), indicating it is a highly effective and usable system. The VR-CBT participants provided an average Igroup Presence Questionnaire score of 42.69 (10.40). There were 24 participants who reported one or more symptoms of transient cybersickness. According to the user experience results, there is a greater decrease in acrophobia symptoms when the usability of the app is high and there is a greater sense of presence. Most participants (n = 47; 72.3%) progressed through all five levels and on average, participants practiced each level twice and the maximum number of sessions spent on a level was 11. demonstrating that the app was engaging and could motivate repeated usage of VR self-guided CBT. | 39/96 (40.6) | Yes |
| Farahimanesh et al. (2023) | - | - | 0/30 (0) | - |
| Kahlon et al. (2023a; 2023b) | - | - | VRET: 4/20 (20);<br><br>VRET + exposure: 4/20 (20) | Yes |
| Khaleghi et al. (2024) | 6-item adapted preference questionnaires to gather user preferences. Semi-structured interviews and a heuristic questionnaire were used to examine playability and usability. | Nine out of ten participants (90%) found the VRET game was easy to learn. The simplicity of using a single button with VR glasses was advantageous for individuals with mobility disabilities. Eighty percent (8/10) of participants experienced no dizziness during extended gameplay. Two participants played the game 7 and 10 times respectively. | 0/10 (0) | - |
| Lacey et al. (2022) | Fast Motion Sickness Scale (FMS) at weeks 3–5. | After completion of the intervention, participants were asked to complete free-text responses to ‘Have you made changes to things you may have avoided because of the phobia after completing this study?’ Free-text comments were made by 39 participants in the active group at week 6 about behavioural changes as a result of the intervention. Three people reported ‘no changes’, and 16 reported reduction in anxiety without commenting on behaviour | 12/63 (19) | Yes |
| | | changes. Twenty reported some degree of behaviour change. | | |
| Lindner et al. (2020) | Simulator Sickness Questionnaire (SSQ), Gatineau Presence Scale (GPS) | There were no significant correlations between treatment outcomes and user experience measures. The average cybersickness score was low, with minimal variation among participants, suggesting that cybersickness was not a notable concern. The Presence Scale showed a positive score. Scores could range from 0 to 20 and participants on average scored 12.39 (4.62). | 0/25 (0) | - |
| Lindner et al. (2019) | Client Satisfaction Questionnaire 8-item version, Negative Effects Questionnaire (NEQ) | The two treatment groups reported similar levels of treatment satisfaction and a similar numbers of negative effects, with a mean of 4.17 (SD = 3.55) out of a possible 32.The negative effect with the greatest endorsement (56%) was "I felt like I was under more stress" The following statements were also endorsed by 34% and 29% of the sample respectively, "I experienced more anxiety" and "I felt that the treatment did not suit me". | 5/25 (20) | Yes |
| Matsumoto et al. (2021) | - | - | VR: 1/24 (4.2)<br><br>VR + CB: 1/23 (4.3) | - |
| Meyer et al. (2022) | Negative Effect Questionnaire (NEQ), Simulation Sickness Questionnaire (SSQ), final interview to evaluate the feasibility and handling of the self-help protocol. | The participants' feelings of general distress, depression, anxiety, and stress were significantly lower at the 2-week follow-up compared to pre-intervention. Feedback obtained from the post-intervention interview was generally positive. However, several participants expressed concerns about the low display quality and resolution, motion sickness, and discomfort of wearing cardboard glasses. | - | - |
| Miller et al. (2023) | Qualitative post intervention interview, | Overall, participants wanted more game-like qualities and interactivity in both the mobile app and VR experiences of the program. Some participants indicated that the program contained too much passive reading or listening to content.<br><br>A total of 17% (5/30) of participants experienced adverse side effects from the Spark program. All reported issues related to the VR component, including headaches (3 participants), feelings of dissociation (1 participant), and dizziness and eye pain (1 participant).<br><br>Participants developed a therapeutic connection with the digital character that led them through the mobile app. Participants likened the digital character to a role model (e.g., a teacher or therapist), who offered guidance and support, aided their understanding, and helped them maintain a sense of responsibility for their progress.<br><br>Participants provided moderate ratings regarding the helpfulness of VR, its ability to enhance their coping skills and be incorporated into routine. There was | 1/30 (3) | Yes |
| | | moderate willingness to recommend VR to a friend. Parents and guardians provided comparatively higher ratings of the program's ease of incorporation into routine and agreement to recommend to a friend and allow their child to continue using the app. | | |
| Piercy et al. (2012) | Presence Questionnaire (PQ), Self-reports | There was no significant increase in presence from pretreatment to post-treatment. Participants commented that their fear of spiders still exists, but they were better able to deal with spiders. For example, one participant reported, ''I feel more at ease when seeing a spider now.'' Another participant reported, ''This experience was definitely worth the time. I do feel like this was a positive experience. I am still a bit scared of spiders but not as much as before.'' | - | - |
| Premkumar et al. (2021) | - | - | 5/32 (15.6) | - |
| Riches et al. (2024) | Sense of Presence Scale (SPS), Video call-based semi-structured interview about their experience of VR | Participants enjoyed engaging in VR when they were stressed because they found the visual scenes and accompanying audio induced relaxation. In particular, the beach and ocean scenes were described as entertaining and calming. However, participants noted a novelty effect, whereby their interest reduced over time due to the monotony of the VR experience and the limited interaction and gamification features. Participants also noted technical problems, including glitches and poor resolution. However, the overall experience was positive and improved participant sleep. | - | - |
| Riva et al. (2021) | - | - | 0/40 (0) | - |
| Senda et al. (2023) | Descriptive statistics of enrolment, side effects, and participant ratings of likeability | A User Experience Survey was completed on VR visits 2 and 9. Participants would likely use VR in the future to manage new symptoms. One participant stopped VR due to side effects of motion sickness and nausea. Other side effects reported include fatigue, heightened emotions, and possible worsening of symptoms. Six ceased treatment due to a perceived lack of benefit, and 4 stopped due to other reasons.<br>A User Experience Survey was completed on VR visits 2 and 9. Participants indicated that they would likely use VR in the future to manage new symptoms.<br>Four participants (17%) demonstrated a clinical response to VR treatment but did not achieve remission. | 12/24 (50) | - |
| Shin et al. (2021) | Simulator Sickness Questionnaire (SSQ) | Participants filled out the SSQ in each session, with an average score of 12.26 across all participants and sessions (SD = 13.26, median = 6, range = 0 – 48). A significant relationship was observed between the number of sessions attended and reduced SSQ scores (*p* = .03). However, no significant association was found between changes in SSQ scores and changes in any clinical scales assessed in the study. | 13/33 (39.4) | Yes |
| Veling et al. (2021) | Simulator Sickness Questionnaire (SSQ) | No serious side effects were reported. Some participants experienced cybersickness and two discontinued use of VRelax due to dizziness and nausea. However, the mean total SSQ score was higher before using VRelax (48.3, SD = 12.7) and decreased afterwards (43.1, SD = 10.9). Five participants stopped VRelax due to a lack of motivation, or because the intervention did not meet expectations. | Phase 1 VR: 1/25 (4%); Phase 2 VR: 6/24 (25) | Yes |
| Zainal et al. (2021) | Simulator Sickness Questionnaire (SSQ), iGroup Presence Questionnaire (IPQ), Qualitative Feedback, Open-ended questions | Participants reported acceptable presence and low levels of simulator sickness. There were high levels of homework compliance and VRE usability ratings. Eighty five percent of participants mentioned that they would recommend the VRE to others troubled by Social Anxiety Disorder. | 3/26 (11.5) | - |

[a] ITT: intention-to-treat analysis
[b] d: Cohen's *d* effect size
[c] HMD: head mounted display

### 3.7 Quality Assessment Results

In most RCT studies, randomisation was appropriately performed, and participants reportedly adhered to their assigned VR intervention. One study did not specify the procedure used to randomise participants to conditions (Khaleghi et al., 2024). Seven RCT studies either did not report a comparable baseline group analysis, or complete outcome date defined as $\geq$ 80% (Chard et al., 2023; Donker et al., 2019; Donker et al., 2020; Lindner et al., 2019, Kahlon et al., 2023a; Kahlon et al., 2023b and Shin et al., 2021). One study (Shin et al., 2021) reported blinding of outcome assessors, which was applied at pre-test measurements. All of the quantitative non-randomised studies utilised validated standard scales, administered the interventions as intended, and used appropriate statistical analyses. While most of these studies had outcome data defined as > 80% (Lindner et al., 2020; Matsumoto et al., 2021; Premkumar et al., 2021; Riva et al., 2021), others had unspecified or high attrition and dropout rates (Meyer et al., 2022; Piercey et al., 2012; Senda et al., 2023). Strengths of recruitment included use of clear inclusion and exclusion criteria, standardised scales with cut off scores, and use of multiple sources to access participants. The two mixed methods studies (Miller et al., 2022; Riches et al., 2024) utilised appropriate qualitative theory and methods to guide their investigation. However, the quantitative phase of one study (Riches et al., 2024) was underpowered due to a small sample size, preventing pre-post analyses of wellbeing outcomes from being conducted as planned. Non-probability sampling methods, such as snowballing and convenience sampling, were used in some studies (e.g., Khaleghi et al., 2024; Meyer et al., 2020; Piercey et al., 2012; Premkumar et al., 2021).

## 4 Discussion

### 4.1 Effectiveness of Self-Guided VR Therapy for Anxiety

A primary aim of our systematic review was to examine the effectiveness of self-guided VR interventions for anxiety treatment. The findings suggest that self-guided VR interventions are generally effective in reducing anxiety symptoms, as 16 out of 21 studies reported significant improvements on standardised measures. Effect sizes ranged from small to large, could be rapidly procured, and all significant reductions in anxiety were sustained at all follow-up measurements. Improvements were obtained for people with public speaking anxiety, social anxiety disorder, specific phobia, depression, bipolar, a psychotic disorder, and sub-clinical anxiety symptoms. These findings are consistent with previous VR research into anxiety disorders and especially research into social anxiety disorder that relies on graded exposure sessions to reduce anxiety (Jingili et al., 2023; Krzstanek et al., 2021). However, it is unclear if the effectiveness results would be maintained beyond 12 months. There were also few studies that focused on children and adolescents, which limits the generalisability of findings. Nonetheless, the findings indicate that self-guided VR therapy for anxiety can help clinicians provide effective stepped care and overcome common treatment barrier issues, such as worker shortages, geographical distance, and long waitlists.

Conversely, five studies in our review did not find a significant reduction in a measure of anxiety. Chard et al. (2023) reported no significant change with participants who had social anxiety disorder treated with three weeks of exposure sessions. Previous research into therapist guided VR interventions to address social anxiety disorder, found that 8 to 12 sessions are more often associated with a significant reduction in social anxiety disorder symptoms (Krzystanek et al., 2021). Therefore, it is possible that more self-guided VRET sessions may be required to significantly reduce social anxiety disorder. Furthermore, Khaleghi et al. (2024) reported no significant change in cat phobia after sessions of self-guided VR exposure therapy that lasted up to three hours. Research suggests that even one long session of up to 45-180 minutes can be effective for treating phobias (Krzystanek et al., 2021), yet the small sample size ($n$ = 10) in Khaleghi et al. (2024) may have reduced the likelihood of detecting significant effects. Moreover, three of the other studies that did not detect a significant decrease in anxiety also had small sample sizes (Chard et al., 2023; Matsumoto et al., 2021; Senda et al., 2023). Also, three of the five studies delivered virtual environments that focused on pleasant environments, relaxation, or mindfulness (Matsumoto et al., 2021; Riva et al., 2021; Senda et al., 2023). In contrast, anxiety reductions were observed in studies that used active VR for exposure therapy. For instance, 11 of 13 (85%) VRET studies reported significant reductions compared to 6 of 8 (75%) relaxation VR intervention studies. This highlights for clinicians a key effectiveness difference between self-guided VR therapy interventions for anxiety that are active (e.g., interactive exposure) versus passive (e.g., watching relaxing scenes).

### 4.2 User Experience of Self-Guided VR Therapy for Anxiety

Another primary aim of our systematic review was to examine the user experience of self-guided VR interventions for anxiety. It is encouraging that self-guided VR interventions were generally able to safely procure reductions in anxiety symptoms in unsupervised settings (e.g., participants' homes). There were no reports of major safety problems such as harmful falls or colliding with real-world objects, repetitive strain injuries, or hygiene issues. This could partially be due to studies making use of self-selecting samples of people already familiar with VR use or the screening out of participants with certain conditions. Nevertheless, the absence of major safety problems was found among participants who reported no prior experience with VR (e.g., see Lindner et al., 2020; Shin et al., 2023). It is possible that some participants may have received help from others (e.g., friends or family members familiar with VR) during their VR treatment. However, the findings imply that people may be adequately supported with the provision of safety instructions prior to VR use or within the VR software itself, such as the removal of dangerous objects and what to do if experiencing simulator sickness (Donker et al., 2020; Lindner et al., 2020). Interestingly, some participants described that while the program was self-guided, they felt like the digital character that guided them through the VR treatment was watching over them (Miller et al., 2023).

Notably, there were limited reports of simulator sickness. A few studies reported participants stopping the use of VR due to side effects, such as motion sickness and nausea (Senda et al., 2023; Veling et al., 2021). Therefore, self-guided VR therapy is not appropriate for all anxious clients. However, Shin et al. (2021) found that lower SSQ scores were significantly associated with the number of VR sessions participants performed. Therefore, it is possible that participants may acclimatise to the VR experience with repeated use and persevere through simulator sickness to achieve reduced anxiety. Importantly, no significant deterioration in anxiety symptoms were reported in any reviewed study from pre- to post-test, despite the potential conflation between anxiety and simulator sickness sharing physiological markers, such as dizziness and nausea (Senda et al., 2023; Veling et al., 2021). This implies that any simulator sickness symptoms are short-lived and will not significantly worsen users' anxiety levels after discontinuing VR use. Overall, the observed safety findings across the included studies in our review evince that self-guide VR therapy for anxiety can be a benign treatment modality for most clients who opt for this treatment approach.

Regarding usability, participants reported that VR headsets were user friendly (Donker et al., 2019) and that learning to use VR technology was remarkably easy (Khaleghi et al., 2024). Usability feedback tended to be more positive in VR interventions that actively focused on graded exposure to treat social anxiety disorder, public speaking anxiety, or a specific phobia. Conversely, the usability feedback tended to be more negative in passive VR interventions that focused on relaxing scenes, pleasant images, or mindfulness. Participants were more likely to stop VR interventions due to the repetitiveness or monotony of the passive virtual environments (Riches et al., 2024), and others provided feedback that the VR environments contained too much passive reading or listening to content, and that it needed to be more interactive (Miller et al., 2023). Furthermore, participants who used the inexpensive cardboard VR headsets were more likely to complain of usability issues due to discomfort and lack of immersion from the cardboard headsets (Chard et al., 2023), low display quality and resolution (Meyer et al., 2022), and technical glitches that hindered participants' experience (Riches et al., 2024). Therefore, clinicians recommending self-guided VR therapy to their anxious clients are likely to obtain better usability feedback if the intervention involves higher quality HMDs and interactive treatment elements in the virtual environments.

Our review findings indicate that anxious people who complete self-guided VR interventions tend to find them acceptable. Many participants reported they would personally use VR interventions in the future for any anxiety issues, and they would recommend VR interventions to other people who struggled with a similar issue (Donker et al., 2020; Lindner et al., 2020; Meyer et al., 2022; Zainal et al., 2021). However, measures of acceptability were often elicited post-treatment after participants successfully completed the treatment protocol. It is probable that attrition occurred for some people because the VR interventions were perceived to be unacceptable (e.g., lacked effectiveness, poor usability, or were deemed unsafe). Therefore, the results on acceptability may be subject to survivorship bias because they did not capture acceptability feedback from participants who dropped out during the treatment.

Self-guided VR therapy for anxiety appears to be an engaging treatment modality. The average attrition rate observed in our review was lower than self-guided internet interventions for anxiety, such as people receiving therapy through websites (Donker et al., 2013). Nonetheless, a higher attrition rate for self-guided VR treatment compared to therapist-guided VR treatment was found (Lindner et al., 2019). Therefore, it is likely rapport with a therapist may assist treatment adherence and reduce dropout rates (Horigome et al., 2020). Also, Senda et al. (2023) reported the highest rate of attrition from pre- to post-test (50%) over a 2-week treatment period. However, their participants had a combination of depression, anxiety, PTSD, and chronic pain conditions. Previous research found that VRET was less effective in reducing anxiety when users had co-morbid illnesses to manage (Krzystanek et al., 2021).

**4.6 Future Research Directions**

There are many future research opportunities for self-guided VR therapy for anxiety. To enhance user engagement, future studies can consider including and examining interactive gamification elements (e.g., puzzles, challenges, interactive scenarios, digital points or badge rewards for completing tasks, social interaction where users can compete with others or share their progress with friends or family, and personalisation whereby users can customise avatars and environments Jingili et al., 2023). Comparative research between self-guided VR interventions and other types of self-help (e.g., internet interventions, bibliography) would help clarify for clinicians which would be most effective for providing stepped care to their anxious clients (Haug et al., 2012). Longitudinal research beyond 12 months can help determine the need and potential effectiveness of VR booster treatments. There were also a lack of studies examining the use of self-guided VR therapy for anxiety within an Australian context. Whilst survey research suggests most Australian mental health practitioners are aware of VR technology (91%), fewer are familiar with applications to clinical care (40%; Chung et al., 2022). VR interventions have high ratings of acceptability (84%) and appropriateness (69%), but lower ratings of feasibility (59%), with concern that those without technological knowledge may show low interest and engagement in treatment (Chung et al., 2022). Therefore, further research is required to examine practitioner perceptions, knowledge, familiarity, and attitudes towards self-guided VR therapy for anxiety, as well as directly test concerns raised about their use within the Australian mental health workforce.

The review found one study that developed a VR program suitable for the treatment of multiple phobias (Lacey et al., 2023). Further research is necessary in this area, as 75% of people with specific phobia are afraid of several objects or situations, with an average of three feared stimuli triggering distress and active avoidance (APA, 2022; Stinson, 2007). In a similar vein, studies tended to exclude individuals with comorbidities, so little is known about how to adapt self-guided VRET programs to suit complex cases. Since self-guided VRET has demonstrated effectiveness for treating social anxiety disorder and specific phobias, future research should also target groups at elevated risk of these conditions. For instance, autism spectrum disorder frequently co-occurs with social anxiety (17%) and specific phobia (30%; APA, 2022); therefore, research should investigate the effectiveness and user experience of self-guided VR programs for anxious people with autism. Further research should also target young people, as social anxiety disorder and most specific phobias have an onset during early development; and phobias a less likely to resolve if they continue into adulthood (APA, 2022). While Miller et al. (2022) found parents agreed that self-led VR could be incorporated into their child's routine, little is known about the nature or extent of the parents' involvement in supporting their child or adolescent's engagement in VR without therapist guidance.

### 4.7 Strengths and Limitations of the Review

This systematic review consolidates available peer-reviewed journal evidence on the use of self-guided VR therapy for anxiety. It has a structured and replicable framework by adopting the Preferred Reporting Items for Systematic Reviews and Meta-Analyses Extension for Systematic Reviews checklist (Page et al., 2021). Our review utilised a broad and comprehensive search strategy that involved multiple reviewers, which improves the validity and reliability of the observed findings. The protocol of this systematic review was prospectively registered online with PROSPERO to minimize selective outcome reporting bias. Nonetheless, this review has several limitations. The limit of English-language studies may have excluded relevant articles published in other languages. The limit of peer-review journal articles may have eliminated other types of literature such as conference presentations, and dissertations. It is also possible that the initial screening of articles based on the title and abstract may have inadvertently eliminated pertinent articles if the keywords and information provided did not clearly relate to self-guided VR for anxiety.

## 5 Conclusion

Self-guided VR therapy for anxiety can reduce symptoms across multiple conditions. Small-to-large treatment effects were observed that could last beyond treatment cessation. Interactive VRET tended to have improved effectiveness and user experience outcomes over passive VR experiences (e.g., watching relaxing environments). A lack of effectiveness was generally observed in studies with passive VR experiences, small sample sizes, and short treatment lengths. Whilst there were limited reports of simulator sickness, self-guided VR therapy was generally found to be engaging, safe, usable and acceptable to anxious people. Future research can help aid clinical decision making and treatment recommendations in this domain by recruiting larger sample sizes, targeting underrepresented anxiety disorder populations, iterating on existing applications to improve their interactive features, and comparing self-guided VR therapy against other established treatments (e.g., therapist guided, internet interventions, bibliotherapy). Overall, our review findings support the potential use of this treatment modality as a form of stepped care when therapist support is limited or unavailable.

## Summary Table

What was already known about the topic.

- Virtual reality (VR) therapy is an effective treatment for anxiety disorders, particularly when guided by a therapist, with significant symptom reduction.
- VR interventions are generally well-accepted by users, with high usability and low risks of adverse effects such as simulator sickness.

What this study adds to our knowledge.

- Demonstrates that self-guided VR therapy can effectively reduce anxiety symptoms with effect sizes ranging from small to large, even without therapist involvement.
- Highlights the moderate to high usability and acceptability of self-guided VR therapy, offering a cost-effective and accessible solution, especially for remote or underserved populations.
- Identifies engaging features, such as gamification and virtual therapists, that enhance user experience and support therapy adherence.
- Addresses gaps by emphasizing the need for research on self-guided VR therapy in underrepresented anxiety populations, particularly within the Australian context.
- Examines safety considerations, including simulator sickness and the ability to manage psychological distress without therapist support.

## Declarations

### Author Contributions

**Winona Graham**: Data curation, Formal analysis, Investigation, Validation, Writing – review & editing. **Russell Drinkwater**: Data curation, Formal Analysis, Investigation, Writing – original draft. **Joshua Kelson**: Conceptualisation, Methodology, Writing - Review & Editing, Supervision, Project administration. **Muhammad Ashad Kabir**: Validation, Writing - Review & Editing.

### Data Availability Statement

This is a review article. Data used for the findings have already been reported in the manuscript itself.

### Funding

Not applicable.

### Conflict of Interest

The authors declare no conflicts of interest.

### References

American Psychiatric Association. (2022). *Diagnostic and statistical manual of mental disorders* (5th ed., text rev.). American Psychiatric Association Publishing.

Andersen, N.J., Schwartzman, D., Martinez, C., Cormier, G., & Drapeau, M. (2023). Virtual reality intervention for the treatment of anxiety disorders: A scoping review. *Journal of Behavior Therapy and Experimental Psychiatry, 81*, Article 101851. https://doi.org/10.1016/j.jbtep.2023.101851

Baghaei, N., Chitale, V., Hlasnik, A., Stemmet, L., Liang, H.-N., & Porter, R. (2021). Virtual reality for supporting the treatment of depression and anxiety: Scoping review. *JMIR Mnetal Health, 8*(9), Article e29681. https://doi.org/10.2196/29681

Balk, S. A., Bertola, D. B., & Inman, V. W. (2013, June). Simulator sickness questionnaire: twenty years later. In *Driving Assessment Conference* (Vol. 7, No. 2013). University of Iowa. https://doi.org/10.17077/drivingassessment.1498

Bandelow, B., Michaelis, S., & Wedekind, D. (2017). Treatment of anxiety disorders. *Dialogues in Clinical Neuroscience, 19*(2), 93-107. https://doi.org/10.31887/DCNS.2017.19.2/bbandelow

Benbow, A.A., & Anderson, P.L. (2019). A meta-analytic examination of attrition in virtual reality exposure therapy for anxiety disorders. *Journal of Anxiety Disorders, 61*, 18-26. https://doi.org/10.1016/j.janxdis.2018.06.006

Caponnetto, P., Triscari, S., Maglia, M., & Quattropani, M. C. (2021). The simulation game—virtual reality therapy for the treatment of Social Anxiety Disorder: A systematic review. *International Journal of Environmental Research and Public Health, 18*(24), 13209. https://doi.org/10.3390/ijerph182413209

Carl, E., Stein, A.T., Levihn-Coon, A., Pogue, J.R., Rothbaum, B., Emmelkamp, P., Asmundson, G.J.G., Carlbring, P., & Powers, M.B. (2019). Virtual reality exposure therapy for anxiety and related disorders: A meta-analysis of randomized controlled trials. *Journal of Anxiety Disorders, 61*, 27-36. https://doi.org/10.1016/j.janxdis.2018.08.003

Chard, I., Van Zalk, N., & Picinali, L. (2023). Virtual reality exposure therapy for reducing social anxiety in stuttering: A randomized controlled pilot trial. *Frontiers in Digital Health*, *5*. https://doi.org/10.3389/fdgth.2023.1061323

Chung, O. S., Johnson, A. M., Dowling, N. L., Robinson, T., Ng, C. H., Yücel, M., & Segrave, R. A. (2022). Are Australian mental health services ready for therapeutic virtual reality? An investigation of knowledge, attitudes, implementation barriers and enablers. *Frontiers in Psychology, 13*, 792663. https://doi.org/10.3389/fpsyt.2022.792663

Clemens, N.A. (2010). Dependency on the psychotherapist. *Journal of Psychiatric Practice, 16*(1), 50-53. https://doi.org/10.1097/01.pra.0000367778.34130.4a

Cullen, A.J., Dowling, N.L., Segrave, R., Morrow, J., Carter, A., Yücel, M. (2021). Considerations and practical protocols for using virtual reality in psychological research and practice, as evidenced through exposure-based therapy. *Behavior Research Methods, 53*, 2725-2742. https://doi.org/10.3758/s13428-021-01543-3

Dhunnoo, P., Wetzlmair, L.-C., & O'Carroll, V. (2024). Extended Reality Therapies for Anxiety Disorders: A Systematic Review of Patients' and Healthcare Professionals' Perspectives. *Sci*, *6*(2), 19. https://doi.org/10.3390/sci6020019

Donker, T., Cornelisz, I., van Klaveren, C., van Straten, A., Carlbring, P., Cuijpers, P., & van Gelder, J.-L. (2019). Effectiveness of self-guided app-based virtual reality cognitive behavior therapy for acrophobia: A randomized clinical trial. *JAMA Psychiatry*, *76*(7), 682. https://doi.org/10.1001/jamapsychiatry.2019.0219

Donker, T., Van Esveld, S., Fischer, N., & Van Straten, A. (2018). 0Phobia – towards a virtual cure for acrophobia: Study protocol for a randomized controlled trial. *Trials*, *19*(1). https://doi.org/10.1186/s13063-018-2704-6

Donker, T., van Klaveren, C., Cornelisz, I., Kok, R. N., & van Gelder, J.-L. (2020). Analysis of usage data from a self-guided app-based virtual reality cognitive behavior therapy for acrophobia: A randomized controlled trial. *Journal of Clinical Medicine*, *9*(6), 1614. https://doi.org/10.3390/jcm9061614

Emmelkamp, P. M. G., & Meyerbröker, K. (2021). Virtual reality therapy in mental health. *Annual Review of Clinical Psychology*, *17*(1), 495–519. https://doi.org/10.1146/annurev-clinpsy-081219-115923

Farahimanesh, S., Serino, S., Tuena, C., Di Lernia, D., Wiederhold, B. K., Bernardelli, L., Riva, G., & Moradi, A. (2023). Effectiveness of a virtual-reality-based self-help intervention for lowering the psychological burden during the covid-19 pandemic: Results from a randomized controlled trial in Iran. *Journal of Clinical Medicine*, *12*(5), 2006. https://doi.org/10.3390/jcm12052006

Goldsworthy, A., Olsen, M., Koh, A., Demaneuf, T., Singh, G., Almheiri, R., Chapman, B., Almazrouei, S., Ghemrawi, R., Senok, A., McKirdy, S., Alghafri, R., & Tajouri, L. (2024). Extended reality head-mounted displays are likely to pose a significant risk in medical settings while current classification remains as non-critical. *Microorganisms, 12*, Article 815. https://doi.org/10.3390/microorganisms12040815

Haug, T., Nordgreen, T., Öst, L.G., & Havik, O.E. (2012). Self-help treatment of anxiety disorders: A meta-analysis and meta-regression of effects and potential moderators. *Clinical Psychology Review*, *32*(5), 425-445. https://doi.org/10.1016/j.cpr.2012.04.002

Hong, Q. N., Gonzalez-Reyes, A., & Pluye, P. (2018). Improving the usefulness of a tool for appraising the quality of qualitative, quantitative and mixed methods studies, the mixed methods appraisal tool (mmat). *Journal of Evaluation in Clinical Practice*, *24*(3), 459–467. https://doi.org/10.1111/jep.12884

Horigome, T., Kurokawa, S., Sawada, K., Kudo, S., Shiga, K., Mimura, M., & Kishimoto, T. (2020). Virtual reality exposure therapy for Social Anxiety Disorder: A systematic review and meta-analysis. *Psychological Medicine*, *50*(15), 2487–2497. https://doi.org/10.1017/s0033291720003785

Jingili, N., Oyelere, S. S., Nyström, M. B., & Anyshchenko, L. (2023). A systematic review on the efficacy of virtual reality and gamification interventions for managing anxiety and depression. *Frontiers in Digital Health*, *5*. https://doi.org/10.3389/fdgth.2023.1239435

Kahlon, S., Gjestad, R., Lindner, P., & Nordgreen, T. (2023). Perfectionism as a predictor of change in digital self-guided interventions for public speaking anxiety in adolescents: A secondary analysis of a four-armed randomized controlled trial. *Cognitive Behaviour Therapy*, *53*(2), 152–170. https://doi.org/10.1080/16506073.2023.2281243

Kahlon, S., Lindner, P., & Nordgreen, T. (2023). Gamified virtual reality exposure therapy for adolescents with public speaking anxiety: A four-armed randomized controlled trial. *Frontiers in Virtual Reality*, *4*. https://doi.org/10.3389/frvir.2023.1240778

Kanwar, A., Malik, S., Prokop, L.J., Sim, L.A., Feldstein, D., Wang, Z., & Murad, M.H. (2013). The association between anxiety disorders and suicidal behaviors: A systematic review and meta-analysis. *Depresson and Anxiety, 30*(10), 917-929. https://doi.org/10.1002/da.22074

Kelson, J.N., Ridout, B., Steinbeck, K., & Campbell, A.J. (2021). The use of virtual reality for managing psychological distress in adolescents: Systematic review. *Cyberpsychology, Behavior, and Social Networking*, *24*(10), 633-641. https://doi.org/10.1089/cyber.2021.0090

Khaleghi, A., Narimani, A., Aghaei, Z., Khorrami Banaraki, A., & Hassani-Abharian, P. (2024). A smartphone-gamified virtual reality exposure therapy augmented with biofeedback for ailurophobia: Development and evaluation study. *JMIR Serious Games*, *12*. https://doi.org/10.2196/34535

Kourtesis, P., Kouklari, E., Roussos, P., Mantas, V., Papanikolaou, K., Skaloumbakas, C., & Pehlivandidis, A. (2023). Virtual reality training of social skills in adults with autism spectrum disorder: An examination of acceptability, usability, user experience, social skills and executive functions. *Behavioural Sciences, 13*(4), 336-334. https://doi.org/10.3390/bs13040336

Krzystanek, M., Surma, S., Stokrocka, M., Romańczyk, M., Przybyło, J., Krzystanek, N., & Borkowski, M. (2021). Tips for effective implementation of virtual reality exposure therapy in phobias—a systematic review. *Frontiers in Psychiatry, 12*. https://doi.org/10.3389/fpsyt.2021.737351

Lacey, C., Frampton, C., & Beaglehole, B. (2022). OVRcome – self-guided virtual reality for specific phobias: A randomised controlled trial. *Australian & New Zealand Journal of Psychiatry*, *57*(5), 736–744. https://doi.org/10.1177/00048674221110779

Lindner, P., Miloff, A., Bergman, C., Andersson, G., Hamilton, W., & Carlbring, P. (2020). Gamified, automated virtual reality exposure therapy for fear of spiders: A single-subject trial under simulated real-world conditions. *Frontiers in Psychiatry*, *11*. https://doi.org/10.3389/fpsyt.2020.00116

Lindner, P., Miloff, A., Fagernäs, S., Andersen, J., Sigeman, M., Andersson, G., Furmark, T., & Carlbring, P. (2019). Therapist-led and self-led one-session virtual reality exposure therapy for public speaking anxiety with consumer hardware and software: A randomized controlled trial. *Journal of Anxiety Disorders*, *61*, 45–54. https://doi.org/10.1016/j.janxdis.2018.07.003

Matsumoto, A., Kamita, T., Tawaratsumida, Y., Nakamura, A., Fukuchimoto, H., Mitamura, Y., Suzuki, H., Munakata, T., & Inoue, T. (2021). Combined use of virtual reality and a chatbot reduces emotional stress more than using them separately. *JUCS - Journal of Universal Computer Science*, *27*(12), 1371–1389. https://doi.org/10.3897/jucs.77237

McMahon, E., & Boeldt, D. (Eds). (2021). *Virtual reality therapy for anxiety: A guide for therapists*. Routledge. https://doi.org/10.4324/9781003154068

Menzies, R. J., Rogers, S. J., Phillips, A. M., Chiarovano, E., de Waele, C., Verstraten, F. A., & MacDougall, H. (2016). An objective measure for the visual fidelity of virtual reality and the risks of falls in a virtual environment. *Virtual Reality*, *20*(3), 173–181. https://doi.org/10.1007/s10055-016-0288-6

Meyer, M. L., Kaesler, A., Wolffgramm, S., Perić, N. L., Bunjaku, G., Dickmann, L., Serino, S., Di Lernia, D., Tuena, C., Bernardelli, L., Pedroli, E., Wiederhold, B. K., Riva, G., & Shiban, Y. (2022). Covid feel good: Evaluation of a self-help protocol to overcome the psychological burden of the covid-19 pandemic in a German sample. *Journal of Clinical Medicine*, *11*(8), 2080. https://doi.org/10.3390/jcm11082080

Miller, I., Peake, E., Strauss, G., Vierra, E., Koepsell, X., Shalchi, B., Padmanabhan, A., & Lake, J. (2023). Self-guided digital intervention for depression in adolescents: Feasibility and preliminary efficacy study. *JMIR Formative Research*, *7*. https://doi.org/10.2196/43260

Mittal, P., Bhadania, M., Tondak, N., Ajmera, P., Yadav, S., Kukreti, A., Kalra, S., & Ajmera, P. (2024). Effect of immersive virtual reality-based training on cognitive, social and emotional skills in children and adolescents with autism spectrum disorder: A meta-analysis of randomised controlled trials. *Research in Developmental Disabilities*, *151*, 104771. https://doi.org/10.1016/j.ridd.2024.104771

Nielsen, J. (2012, January 3). *Usability 101: Introduction to usability.* Nielsen Norman Group. Retrieved April 18, 2024, from https://www.nngroup.com/articles/usability-101-introduction-to-usability/

Oing, T., & Prescott, J. (2018). Implementations of Virtual Reality for Anxiety-Related Disorders: Systematic Review. *JMIR serious games*, *6*(4), e10965. https://doi.org/10.2196/10965

Page, M.J., McKenzie, J.E., Bossuyt, P.M., Boutron, I., Hoffmann, T.C., Mulrow, C.D., Shamseer, L., Tetzlaff, J.M., & Moher, D. (2021). Updating guidance for reporting systematic reviews: Development of the PRISMA 2020 statement. *Journal of Clinical Epidemiology*, *134*, 103-112. https://doi.org/10.1016/j.jclinepi.2021.02.003

Pallavicini, F., Orena, E., Achille, F., Cassa, M., Vuolato, C., Stefanini, S., Caragnano, C., Pepe, A., Veronese, G., Ranieri, P., Fascendini, S., Defanti, C.A., Clerici, M., & Mantovani, F. (2022). Psychoeducation on stress and anxiety using virtual reality: A mixed-methods study. *Applied Sciences*, *12*, Article 9110. https://doi.org/10.3390/app12189110

Pellas, N., Dengel, A., & Christopoulos, A. (2020). A scoping review of immersive virtual reality in STEM education. *IEEE Transactions on Learning Technologies*, *13*(4), 748-761. https://doi.org/10.1109/TLT.2020.3019405

Pelucio, L., Quagliato, L. A., & Nardi, A. E. (2024). Therapist-Guided Versus Self-Guided Cognitive-Behavioral Therapy: A Systematic Review. *The primary care companion for CNS disorders*, *26*(2), 23r03566. https://doi.org/10.4088/PCC.23r03566

Piercey, C. D., Charlton, K., & Callewaert, C. (2012). Reducing anxiety using self-help virtual reality cognitive behavioral therapy. *Games for Health Journal*, *1*(2), 124–128. https://doi.org/10.1089/g4h.2012.0008

Porter, R. J., Beaglehole, B., & Baghaei, N. (2023). Virtual Reality Technology in the treatment of anxiety – progress and future challenges. *Expert Review of Neurotherapeutics*, *23*(12), 1047–1049. https://doi.org/10.1080/14737175.2023.2289574

Premkumar, P., Heym, N., Brown, D. J., Battersby, S., Sumich, A., Huntington, B., Daly, R., & Zysk, E. (2021). The effectiveness of self-guided virtual-reality exposure therapy for public-speaking anxiety. *Frontiers in Psychiatry*, *12*. https://doi.org/10.3389/fpsyt.2021.694610

Riches, S., Kaleva, I., Nicholson, S. L., Payne-Gill, J., Steer, N., Azevedo, L., Vasile, R., Rumball, F., Fisher, H. L., Veling, W., & Valmaggia, L. (2024). Virtual reality relaxation for stress in young adults: A remotely delivered pilot study in participants' homes. *Journal of Technology in Behavioral Science*. https://doi.org/10.1007/s41347-024-00394-x

Riva, G., Bernardelli, L., Castelnuovo, G., Di Lernia, D., Tuena, C., Clementi, A., Pedroli, E., Malighetti, C., Sforza, F., Wiederhold, B. K., & Serino, S. (2021). A virtual reality-based self-help intervention for dealing with the psychological distress associated with the COVID-19 lockdown: An effectiveness study with a two-week follow-up. *International Journal of*

*Environmental Research and Public Health*, *18*(15), 8188. https://doi.org/10.3390/ijerph18158188

Rowland, D. P., Casey, L. M., Ganapathy, A., Cassimatis, M., & Clough, B. A. (2022). A decade in review: A systematic review of virtual reality interventions for emotional disorders. *Psychosocial Intervention*, *31*(1), 1–20. https://doi.org/10.5093/pi2021a8

Schröder, D., Wrona, K. J., Müller, F., Heinemann, S., Fischer, F., & Dockweiler, C. (2023). Impact of virtual reality applications in the treatment of anxiety disorders: A systematic review and meta-analysis of randomized-controlled trials. *Journal of Behavior Therapy and Experimental Psychiatry*, *81*, Article 101893. https://doi.org/10.1016/j.jbtep.2023.101893

Senda, M. C., Johnson, K. A., Taylor, I. M., Jensen, M. M., Hou, Y., & Kozel, F. A. (2023). A pilot trial of stepwise implementation of virtual reality mindfulness and accelerated transcranial magnetic stimulation treatments for dysphoria in neuropsychiatric disorders. *Depression and Anxiety*, *2023*, 1–16. https://doi.org/10.1155/2023/9025984

Shahid, S., Kelson, J., & Saliba, A. (2024). Effectiveness and user experience of virtual reality for Social Anxiety Disorder: Systematic Review. *JMIR Mental Health*, *11*, Article e48916. https://doi.org/10.2196/48916

Shin, B., Oh, J., Kim, B.-H., Kim, H. E., Kim, H., Kim, S., & Kim, J.-J. (2021). Effectiveness of self-guided virtual reality–based cognitive behavioral therapy for panic disorder: Randomized controlled trial. *JMIR Mental Health*, *8*(11), Article e30590. https://doi.org/10.2196/30590

Siddaway, A.P., Wood, A.M., & Hedges, L.V. (2019). How to do a systematic review: a best practice guide for conducting and reporting narrative reviews, meta-analyses, and meta-syntheses. *Annual Review of Psychology*, *70*(1), 747-770. https://doi.org/10.1146/annurev-psych-010418-102803

Veling, W., Lestestuiver, B., Jongma, M., Hoenders, H. J., & van Driel, C. (2021). Virtual reality relaxation for patients with a psychiatric disorder: Crossover randomized controlled trial. *Journal of Medical Internet Research*, *23*(1). https://doi.org/10.2196/17233

Wang, Z., & Chan, M.-T. (2024). A systematic review of Google Cardboard used in Education. *Computers & Education: X Reality*, *4*, Article 100046. https://doi.org/10.1016/j.cexr.2023.100046

Wang, Y., Guo, L., & Xiong, X. (2022). Effects of virtual reality-based distraction of pain, fear and anxiety during needle-related procedures in children and adolescents. *Frontiers in Psychology*, *13*, Article 842847. https://doi.org/10.3389/fpsyg.2022.842847

Wiebe, A., Kannen, K., Selaskowski, B., Mehren, A., Thöne, A. K., Pramme, L., Blumenthal, N., Li, M., Asché, L., Jonas, S., Bey, K., Schulze, M., Steffens, M., Pensel, M. C., Guth, M., Rohlfsen, F., Ekhlas, M., Lügering, H., Fileccia, H., Pakos, J., … Braun, N. (2022). Virtual reality in the diagnostic and therapy for mental disorders: A systematic review. *Clinical psychology review*, *98*, 102213. https://doi.org/10.1016/j.cpr.2022.102213

Wilmer, M.T., Anderson, K., & Reynolds, M. (2021). Correlates of quality of life in anxiety disorders: Review of recent research. *Current Psychiatry Reports*, *23*, article number 77. https://doi.org/10.1007/s11920-021-01290-4

Wu, J., Sun, Y., Zhang, G., Zhou, Z., & Ren, Z. (2021). Virtual Reality-Assisted Cognitive Behavioral Therapy for Anxiety Disorders: A Systematic Review and Meta-Analysis. *Frontiers in psychiatry*, *12*, 575094. https://doi.org/10.3389/fpsyt.2021.575094

Zainal, N. H., Chan, W. W., Saxena, A. P., Taylor, C. B., & Newman, M. G. (2021). Pilot randomized trial of self-guided virtual reality exposure therapy for Social Anxiety Disorder. *Behaviour Research and Therapy*, *147*, 103984. https://doi.org/10.1016/j.brat.2021.103984